\documentclass[runningheads]{llncs}
\usepackage[T1]{fontenc}
\usepackage{graphicx}
\usepackage{graphicx,amssymb}
\usepackage[linesnumbered,ruled,vlined]{algorithm2e}
\usepackage{amsmath}
\usepackage[usenames]{color}
\usepackage{longtable}
\usepackage{booktabs}
\usepackage{url}
\usepackage{mathrsfs}
\usepackage{amsfonts}

\usepackage{enumerate}
\usepackage{multirow}
\usepackage{verbatim}
\usepackage[pagewise]{lineno}
\usepackage{makecell}
\usepackage{subfigure}
\usepackage{threeparttable}

\usepackage{diagbox}
\usepackage{marvosym}

\begin{document}
\title{A Partition-and-Merge Algorithm for Solving the Steiner Tree Problem in Large Graphs}
\titlerunning{A Partition-and-Merge Algorithm for the Steiner Tree Problem}
%

\author{Ming Sun\inst{1} \and
Xinyu Wu\inst{1} \and
Yi Zhou\inst{1}\thanks{Corresponding author.} \and 
Jin-Kao Hao\inst{2} \and
Zhang-Hua Fu \inst{3,4}}

\authorrunning{M. Sun et al.}

\institute{University of Electronic Science and Technology of China, China \\ 
\email{sunm@uestc.edu.cn}, \email{wu.xinyu@outlook.com}, \email{zhou.yi@uestc.edu.cn} 
\and
LERIA, Universit$\acute{e}$ d'Angers, 2 Boulevard Lavoisier, 49045 Angers, France \\
\email{jin-kao.hao@univ-angers.fr} 
\and
Shenzhen Institute of Artificial Intelligence and Robotics for Society, China
\and
The Chinese University of Hong Kong, Shenzhen, China \\
\email{fuzhanghua@cuhk.edu.cn} \\
}

\maketitle              
\begin{abstract}
The Steiner tree problem aims to determine a minimum edge-weighted tree that spans a given set of terminal vertices from a given graph.
In the past decade, a considerable number of algorithms have been developed to solve this computationally challenging problem. 
However, existing algorithms typically encounter difficulties for solving large instances, i.e., graphs with a high number of vertices and terminals.
In this paper, we present a novel partition-and-merge algorithm to effectively solve this problem in large graphs. 
The algorithm breaks the input network into small subgraphs and then merges the subgraphs in a bottom-up manner. 
In the merging procedure, partial Steiner trees in the subgraphs are also created and optimized by efficient local optimization. 
When the merging procedure ends, the algorithm terminates and reports the final solution for the input graph.
We evaluated the algorithm on a wide range of benchmark instances, showing that the algorithm outperforms the best-known algorithms on large instances and competes favorably with them on small or medium-sized instances.

\keywords{Steiner tree problem  \and Network Design \and  Local search \and Partition-and-merge\and Large graphs.}
\end{abstract}

\section{Introduction}

The Steiner tree problem (STP) is a fundamental network design problem to determine the topology of optical networks \cite{hsieh2007all}, telephone networks \cite{cheng2013steiner}, or even multimedia networks \cite{kun2006dynamic}.
The input of the classic STP consists of an undirected edge-weighted  graph (or network) $G$ with a vertex set $V(G)$, an edge set $E(G)$, a non-negative weight function on its edges $c_G:E(G)\rightarrow \mathbb{R}^{+}$, and a set of \emph{terminals} $A\subseteq V(G)$.
The problem is to determine a minimum Steiner tree, that is, a tree $T$ spanning all terminals in $A$ (and possibly other vertices) and minimizes $\sum_{e\in E(T)}c_G(e)$.
In fact, this optimization problem has also been studied in many communities, including operations research and theoretical computer science.

The decision version of STP is one of Karp's 21 NP-Complete problems.  In recent decades, a rich number of theoretical studies have been devoted to this problem.
For instance, it is known that STP can be solved in time $O(2^tn^2+nm)$  \cite{bjorklund2007fourier} where $n, m$ and $t$ denote the number of vertices, edges, and terminals, respectively. The current best approximation ratio is 1.39 \cite{byrka2013steiner}, and there is no 1.01 approximation algorithm (unless P = NP) \cite{chlebik2008steiner}.
In practice, except in the area of network design, STP is also modeled in many other fields like computational biology \cite{cheng2004steiner}, VLSI design \cite{held2011combinatorial}, and social network analysis \cite{lappas2009finding}.
Notably, two international competitions, the 11th DIMACS Implementation Challenge in 2014 \cite{johnson201411th} and the 3rd Parameterized Algorithms and Computational Experiments (PACE) Challenge in 2018 \cite{bonnet2018pace} have been dedicated to solving STP practically, which demonstrates the interest of the research community for this challenging problem and its variants such as the directed Steiner tree problem, the prize-collecting Steiner tree problem, etc.

Due to the great importance of STP, there exists a considerable number of both exact and heuristic solution algorithms for the problem.
Exact algorithms typically rely on two search frameworks, dynamic programming and integer linear programming.
To our knowledge, exact algorithms can solve problem instances with up to thousands of vertices or hundreds of terminals in a reasonable time, mostly thanks to some new ILP techniques like vertex formulation and local branching \cite{carrabs2021minimum,fischetti2017thinning}.
However, to deal with larger instances whose optimality cannot be determined by exact algorithms, heuristic algorithms become indispensable \cite{fischetti2017thinning,fu2017swap,pajor2018robust,ribeiro,uchoa2012fast}.
For example, for some VLSI graphs with up to ten thousand vertices, no exact algorithms can produce a feasible solution.
According to the final report of the 11th DIMACS Challenge, PUW \cite{pajor2018robust} and Staynerd \cite{fischetti2017thinning} are considered the best global heuristics compared to other competing algorithms (\emph{AB} \cite{althaus2014algorithms}, mozartballs \cite{fischetti2017thinning}, polito \cite{biazzo2012performance}, and scipjack \cite{gamrath2017scip}) in the heuristic track.
Interestingly, the exact algorithm mozartballs, which also uses some search techniques of Staynerd, is the winner in the exact track during the 11th DIMACS Challenge. As for the more recent PACE Competition, an evolutionary algorithm developed by the CIMAT Team ranked first in the heuristic track \cite{bonnet2018pace}.
Considering that PUW and Staynerd were absent from PACE, we consider that PUW, Staynerd and CIMAT are among the most competitive heuristics for STP so far. Notably, all these heuristic algorithms use local search as their key optimization component.

Today, large STP instances with at least several thousand vertices appear ubiquitously in many applications. For example, the well-known GEO instances have more than 100,000 vertices or up to 5,000 terminals.
We observed that existing heuristic algorithms, including the aforementioned DIMACS and PACE winners, attain their limit when they are applied to such large instances, leading to large gaps to the best lower bounds. In this work, our aim is to propose a novel heuristic algorithm that is able to better solve large-scale STP instances. The contributions can be summarized as follows.

\begin{itemize}
   \item We design a partition-and-merge (PM) algorithm to deal with the large-scale nature of the given instance. 
   PM first breaks the input graph into a number of smaller subgraphs.
   Then it finds partial solutions for these subgraphs and pairwise merges the subgraphs and solutions at the same time.
   The algorithm provides a high-quality complete solution when all the subgraphs are merged into the input graph.
   Meanwhile, we investigate different merging heuristics. 
   \item We conduct extensive experiments, with a focus on very large instances, to evaluate the PM algorithm and different techniques. 
   We show that our algorithm is very competitive in solving large and hard instances. Remarkably, for the largest publicly available STP instances, the EFST family, PM outperforms the known heuristic solvers including the winners of the 11th DIMACS and 3rd PACE competitions.
\end{itemize}

\section{Basic Notations}\label{basic notations}
A STP instance is composed of an edge-weighted undirected graph $G$ and a subset of terminal vertex $A$.
We use $V(G)$ to denote the vertex set, $E(G)$ to denote the edge set, and $c_G:E(G)\rightarrow \mathbb{R}^{+}$ to denote the edge-weight function of $G$.
So $A\subseteq V(G)$.
When the context is clear, we use $n$, $m$ and $t$ to denote the number of vertices, edges, and terminals, respectively.
For two arbitrary vertices $u,v\in V(G)$, the \emph{distance} $d_G(u,v)$ between them is the smallest total weight among the paths between $u$ and $v$ in $G$.
For a vertex subset $S\subseteq V(G)$, $G[S]$ denotes the subgraph induced by $S$.

A \emph{partition} of the graph $G$, $\mathcal{P}=\{H_1, ..., H_{|\mathcal{P}|} \}$,  is a set of vertex-induced subgraphs of $G$ such that $V(H_i)\cap V(H_j)= \emptyset$ for any $1\le i<j \le |\mathcal{P}|$ and $\bigcup_{1\le i\le |\mathcal{P}|}{V(H_i)}=V(G)$.
$(u,v)\in E(G)$ is a \emph{cutting edge} of $\mathcal{P}$ if $u$ and $v$ are in different subgraphs of $\mathcal{P}$.

Give an input instance with graph $G$ and terminal set $A\subseteq V(G)$, a \emph{Voronoi diagram} $\mathcal{D}=\{G_1,...,G_t\}$ is a special partition of $G$ such that each subgraph $G_i$ contains exactly one terminal, denoted by $a_i$, and for all other vertex $u\in V(G_i)\setminus\{a_i\}$, $a_i$ is the nearest terminal of $u$ in $G$, that is, $d_G(u,a_i)=min_{a\in A}d_G(u,a)$. Clearly, the number of subgraphs in a Voronoi diagram is exactly $t$.
For an undirected graph $G=(V,E)$, the Voronoi diagram can be computed by Dijkstra's shortest path algorithm in $O(m + n \log n)$ time (with ties breaking randomly) \cite{uchoa2012fast}.

\section{The Partition-and-Merge Algorithm}\label{PM Framework}

Our Partition-and-Merge (PM) algorithm is described in Alg. \ref{Framework}.
The input of PM consists of an input graph $G=(V,E)$, a terminal set $A\subseteq V$, two parameters $\rho\le |A|$ and $\gamma>1$, and two maximum iteration numbers $I_{short}$ and $I_{long}$.

PM maintains a partition of $G$, that is, $\mathcal{P}=\{H_1,...,H_{|\mathcal{P}|}\}$. 
As PM runs, the subgraphs in $\mathcal{P}$ are merged in pairwise.
The \emph{merge operation} concatenates $H_i$ and $H_j$ into a larger graph $H'$ by adding the cutting edges between $H_i$ and $H_j$.
In $H'$, the edge weights are the same as their weights in $E(H_i)$ and $E(H_j)$ from $H_i$ and $H_j$, respectively, but the weights of the cutting edges between $H_i$ and $H_j$ are the same as in the original input graph $G$.
It is important to notice that the weight of the edge $e$ in $E(H_i)$ is possibly different from its weight in $E(G)$.
The reason is that PM modifies the weights of the edges of the subgraphs in $\mathcal{P}$ as the search runs.

\begin{algorithm}[htb]
	\footnotesize
	\caption{The Partition-and-Merge Algorithm}\label{Framework}
	\KwIn{Graph $G$, terminal set $A\subseteq V(G)$, parameters $\rho\in [0,|A|]$, $\gamma \in [1,\infty)$, the maximum iteration numbers $I_{short}$ and $I_{long}$ for local search.}
	\KwOut{Steiner tree $T$}
	Compute a Voronoi diagram $\mathcal{P}=\{H_1, ..., H_{|\mathcal{P}|}\}$ with respect to $G$ and $A$. \\
    Initially, $level\gets 1$\\
	\While{$|\mathcal{P}| >  \rho$}{
		$\mathcal{P} \longleftarrow$ OneRoundMerge$(G, \mathcal{P}, \lceil \frac{n\gamma^{level}}{t} \rceil)$\\
		$level \gets level + 1 $\\
	}
	\While{$|\mathcal{P}| > 1$}{
		\For{each $H_i \in \mathcal{P}$}{
			$T_i$ $\longleftarrow$ LocalOptimize($G[V(H_i)]$, $H_i$, $A\cap V(H_i)$, $I_{short}$)\\
			$H_i' \gets$ SolutionStoring($H_i$,$T_i$)\\
            $\mathcal{P}\gets \mathcal{P}\setminus \{H_i\}\cup \{H_i'\}$
		}
		$\mathcal{P} \longleftarrow$ OneRoundMerge$(G, \mathcal{P}, \lceil \frac{n\gamma^{level}}{t} \rceil)$.\\
		$level\gets level + 1$\\
	}

	There is only one graph $H_1$ in $\mathcal{P}$.\\
	$T\longleftarrow$ LocalOptimize($G$, $H_1$, $A$, $I_{long}$)\\
	\Return{$T$}
\end{algorithm}

\subsection{Two-phase Merging Procedure}
Initially, as shown in line 1 of Alg. \ref{Framework}, PM builds $\mathcal{P}$ by computing the Voronoi diagram, so each subgraph in $\mathcal{P}$ contains one terminal vertex.
The algorithm then performs a two-phase merging procedure, which gradually merges the subgraphs of $\mathcal{P}$ into a whole graph.
The parameter $\rho$ is used to set the time to start the second phase and $\gamma$ controls the size gap between different subgraphs in $\mathcal{P}$.

In the first phase (lines 3-5), \emph{OneRoundMerge} is used to pairwise merge the graphs in $\mathcal{P}$.
In general, \emph{OneRoundMerge} selects two subgraphs by a specific heuristic rule and joins them to form a larger graph.
We introduce different merging rules in Section \ref{Merging Heuristics}.
\emph{OneRoundMerge} also makes sure that the subgraph after a round of merge has no more than $\lceil \frac{n\gamma^{level}}{t} \rceil$ vertices (so that the subgraphs in $\mathcal{P}$ have roughly the same scale).
Clearly, $\lceil \frac{n\gamma^{level}}{t} \rceil$ increases geometrically with a factor of $\gamma$ as $level$ increases.
For example, if $\gamma=2$, this value doubles at every level.

The second phase, as in lines 6-12, begins when there are no more than $\rho$ subgraphs in $\mathcal{P}$.
In this phase, for each subgraph $H_i$, we additionally carry out a local search algorithm, LocalOptimize, to find a (high-quality) Steiner tree in $G[V(H_i)]$.
It is worth noting that LocalOptimize finds a starting solution from $H_i$, but improves the solution in $G[V(H_i)]$.
Assume $T_i$ is a Steiner tree obtained by LocalOptimize($G[V(H_i)]$, $H_i$, $A\cap V(H_i)$, $I_{short}$).
$T_i$ spans terminals $A\cap V(H_i)$ in both $H_i$ and $G[V(H_i)]$.
Our intuition is to keep the information that $T_i$ is a high-quality solution of graph $H_i$, so that the further search will benefit from this information.
This is implemented by SolutionStoring method, which essentially changes the edge weights of $H_i$.
We will introduce two different ways for SolutionStoring in Section \ref{Edge-reweighting}.
After local optimization and solution storing for each subgraph, we continue using \emph{OneRoundMerge} with the same heuristic in the first phase to merge the subgraphs.



\textbf{Example.} We consider the example in Fig. \ref{fig:PM2}, which describes the partition-and-merge procedure. We set  $\gamma=2$ and $\rho=6$. The size of each subgraph approximately doubles after each level of merging. The Steiner tree solution is computed when there are no more than 6 subgraphs.

\begin{figure*}[htp]
	\centering
	\includegraphics[width=8cm]{pic/bottom_up.pdf}
	\caption{The bottom-up partition-and-merge procedure. Terminal vertices are marked by black dots. The sub-solutions computed for subgraphs are marked by bold lines.}
	\label{fig:PM2}
\end{figure*}


\subsection{Merging Heuristics}\label{Merging Heuristics}

As mentioned, the \textit{OneRoundMerge} procedure repeatedly selects a pair of subgraphs ($H_i$, $H_j$) from $\mathcal{P}$ and merges them into one subgraph.
In Alg. \ref{Merging procedure}, we show the general procedure of \textit{OneRoundMerge} using the so-called \emph{random rule}. We will also discuss the \emph{random-edge} and \emph{min-distance} rules afterward.



\begin{algorithm}[]
	\footnotesize
	\caption{One round of merging by random heuristic }\label{Merging procedure}
	\KwIn{Graph $G$, a partition $\mathcal{P}$, subgraph size limit $\omega$}
	\KwOut{A new partition of $G$}
	\While{true}{

        \fbox{     \parbox{0.85\textwidth}{
        \textbf{Random rule}:\\
        From all pairs $(H_k, H_l)\in \mathcal{P}\times\mathcal{P}$ such that $|V(H_k)|+|V(H_l)|\le \omega$ and $cut_{\mathcal{P}}(H_k, H_l)\neq \emptyset$,
        randomly pick one $(H_i, H_j)$.
        }} \\
		Merge $H_i$ and $H_j$ to $H'$.\\
        $\mathcal{P}\gets \mathcal{P}\setminus\{H_i, H_j\}\cup\{H'\}$\\
        \If{ none of the pairs of graphs in $\mathcal{P}$ meet the random heuristic} {
            break
        }
	}
	\Return{$\mathcal{P}$}
\end{algorithm}

\subsubsection{Random Rule}
By this rule, we randomly select two subgraphs that are \emph{adjacent}, that is, there is at least one cutting edge between them, and the total number of vertices in the two subgraphs does not exceed $\omega$.
Then we merge the two subgraphs and repeat the procedure until no such pair of graphs exits.
The intuition behind this heuristic is to keep each subgraph in $\mathcal{P}$ connected and balanced in terms of vertex number after merging.

\subsubsection{Random-edge Rule}
 The random-edge rule, as shown in below, resembles the random rule except that it picks a cutting edge randomly rather than a pair of adjacent subgraphs randomly. Clearly, this rule concerns the number of cutting edges between every pair of adjacent subgraphs.
 Using this rule to replace line 3 in Alg. \ref{Merging procedure}, we can obtain another version of \textit{OneRoundMerge} with the random-edge heuristic.

\vspace{1.3ex}

\fbox{ \parbox{0.9\textwidth}{
		\textbf{Random-edge rule}:\\
		From all cutting edges $(x, y)$ such that $x \in V(H_k)$, $y \in V(H_l)$ and $|V(H_k)|+|V(H_l)|\le \omega$,
		randomly pick one edge $(u, v)$.\\
		Suppose $u \in V(H_i)$ and $ v \in V(H_j)$ without loss of generality.
}}\\

We note that this rule is inspired by Karger's minimum-cut randomized algorithm \cite{karger1993global}.
As we know, Karger's algorithm finds the minimum cut of a graph with high probability by repeatedly merging a randomly selected edge.
Thus, it is believed that this heuristic gives preference to the pair of adjacent subgraphs that are \emph{closely connected}, that is, there are more cutting edges between them.


\subsubsection{Min-distance Rule}
Let $A_i=A\cap V(H_i)$ and $A_j=A\cap V(H_j)$ where $A$ is the input terminal set, $H_i$ and $H_j$ are two subgraphs in $\mathcal{P}$.
For each pair of $H_i$ and $H_j$, we compute $d_{G}(A_i,A_j)$, the minimum value over all the shortest lengths of path between any two terminals in $A_i\times A_j$, i.e., $d_G(A_i,A_j)=\min_{(u,v)\in A_i\times A_j} d_G(u,v)$. 
Then, the min-distance is shown as follows. 
Using this rule to replace line 3 in Alg. \ref{Merging procedure}, we obtain \textit{OneRoundMerge} with the min-distance heuristic.

\vspace{1.3ex}

\fbox{     \parbox{0.9\textwidth}{
		\textbf{Min-distance rule}: \\
		Find all pairs $(H_k, H_l)\in \mathcal{P}\times\mathcal{P}$ such that $|V(H_k)|+|V(H_l)|\le \omega$ and $cut_{\mathcal{P}}(H_k, H_l)\neq \emptyset$, pick a pair $(H_i, H_j)$ with smallest $d_G(A_i,A_j)$, ties breaking randomly
}}\\


The running time of the min-distance rule  bounded by $O(|E|)$ due to the early computation of the Voronoi diagram.
For each non-terminal vertex $u\in H_i\subseteq A_i$, we already know its nearest terminal $a_{u}=argmin_{a\in A_i}d_{G}(a,u)$ after computing the Voronoi diagram.
Thus, for each cutting edge $(u,v)\in E_{\mathcal{P}}$, we first obtain $d(u,v)=d_G(u,a_u) + c_G(u, v) +  d_G(v,a_v)$ where $a_u$  and $a_v$ are the nearest terminals of $u$ and $v$, respectively. Then, we randomly pick up an $(u,v)$ of the smallest values $d(u,v)$ and merge its corresponding subgraphs $H_i$ and $H_j$. Because we only iterate over the edges in $E_{\mathcal{P}}$, the above running time follows.

\begin{figure*}[!htp]
	\centering
	\includegraphics[width=8cm]{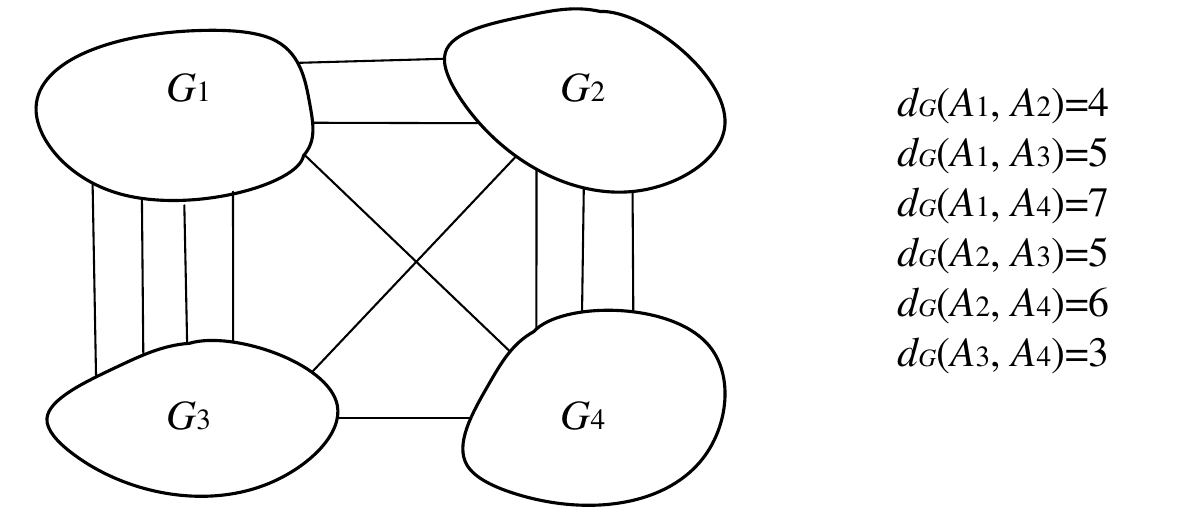}
	\caption{Example for three merging rules. $G_1-G_4$ are four subgraphs. Lines connecting the subgraphs represent the cutting edges between subgraphs. The minimum distance between the terminals of each two subgraphs is shown in the right-side of the figure.}
	\label{fig:Merging}
\end{figure*}

\textbf{Example.}   
In Figure \ref{fig:Merging}, there are currently four subgraphs $G_1,...,G_4$ and any two subgraphs have fewer than $\omega$ vertices. Now, we need to select a pair of adjacent subgraphs to merge. 
If the random rule is used, the probability that each pair of adjacent subgraphs is selected is the same.
If the random-edge rule is used, the probability of selecting subgraphs $(G_1, G_3)$ is the highest, followed by $(G_2, G_4)$, $(G_1,G_2)$ and $(G_3, G_4)$. 
If the min-distance rule is used, $(G_3, G_4)$ is definitely selected for merging.

\subsection{Solution Storing}\label{Edge-reweighting}

The motivation of solution storing is to guide future search by the fact that $T_i$ is a high-quality solution in $H_i$.
To achieve this, we simply change the edge weight of $H_i$ so that we can apply the same local search algorithm.
Here, we introduce a random way to reweight the edges in $H_i$. 

For any edge $e\in E(H_i)$, we score the edge weight $c_{H_i}(e)$ as follows, where $\alpha$ is a random real number from $[0,0.75]$.
\[ c_{H_i}(e) = \left\{
\begin{aligned}
&(1-\alpha) c_{G}(e) & e\in E(T_i) \\
&(1+\alpha) c_{G}(e) & \mbox{otherwise}
\end{aligned}
\right.
\]

Clearly, the weights of edges in $E(T_i)$ are reduced, while the weights of edges not in $E(T_i)$ are increased, implying that the edges of $E(T_i)$ are preferred.

\subsection{Local Optimization}
\label{Local Search Heuristic}

The local optimization procedure, LocalOptimize$(G, H, A, I_{max})$ is a population-based search algorithm modified from the state-of-the-art PUW solver \cite{pajor2018robust}.
The original graph $G$, the edge reweighted graph $H$, terminal set $A$, maximum number of iterations $I_{max}$.

In summary, LocalOptimize follows the scheme of genetic programming. It maintains a pool of elite solutions $\mathcal{Q}$.
In each generation,  it produces new solutions to update the elite pool -- It first builds a new Steiner tree $T$ from $H$ (Note that $H$ is a graph with biased edge weights by solution storing).
Then, $T$ is improved by local search, but the local search runs on the initial graph $G$ without biased edge weights.
Afterward, the improved solution $T$ exchanges information with another solution from the pool by combination, resulting in a new solution $T'$.
Finally, $T$  and $T'$ are used to update the pool.

For detailed description of the local optimization algorithm, please refer to the full version.

\section{Experiments}\label{sec-experiments}

\subsection{Settings}
Our algorithm is programmed in C++ and compiled in g++ with the optimization option '-O3'. 
\footnote{Source codes are publicly available at \url{https://github.com/xyu03/PM.git}.}
All experiments were conducted on a Linux server with Intel Xeon Gold 6130 (2.10 GHz) processor.


\textit{Parameters Setting.}
In the experiments, $\rho$ is set to 64 and $\gamma$ is set to 2.
For all test instances, we set the total time limit $t_{total}=3600$ seconds (1h).
The two parameters for controlling the number of iterations in LocalOptimize, $I_{short}$ and $I_{long}$, are set to 256 and 2048, respectively.
However, with these configurations, it is possible that we may not obtain a complete solution after 3600 s for some graphs.
This happens when there are still more than one subgraph in $\mathcal{P}$ after this time frame.
Thus, we additionally require that each LocalOptimize runs at most $\frac{0.7(\gamma-1)}{\gamma (\rho-1)}\times t_{total}$ seconds for each subgraph.
This implies that at least $0.3\times t_{total}$ seconds are allocated for running LocalOptimize with the final graph, that is, line 14 in Alg. \ref{Framework}.

\subsection{Performance Assessment}



\paragraph{Results on the Set of 113 Large Instances} \label{result_large}

We first test the algorithm in 113 Large Instances \footnote{All instances are downloaded from site \url{http://dimacs11.zib.de/downloads.html}.}. They are grouped in 8 datasets shown in Table \ref{Information for large datasets}, which indicates the number of instances, average vertex number, average edge number, average terminal number, and source of the dataset. GEO, I and EFST contain larger instances (with more vertices, edges, and terminals) compared to ES, VLSI and TSPFST.
\begin{table}[h]
	\centering
	\caption{Information about the 8 large datasets}
	\label{Information for large datasets}
	\setlength{\tabcolsep}{0.9mm}{
	\begin{threeparttable}
		\begin{tabular}{cccccc}
			\hline
			Dataset     & Num & ave.$n$   & ave.$m$   & ave.$t$ & Source                     \\
			\hline
			ES          & 16   & 4245               & 6235               & 1563             & GeoSteiner generator    \\
			TSPFST      & 10   & 4687               & 7043               & 1871             & TSPLIB                                 \\
			VLSI        & 10   & 17926              & 30973              & 452              & VLSI applications                      \\
			GEO         & 22   & 105906             & 150767             & 1845             & Telecommunication networks             \\
			I           & 10   & 65120              & 106450             & 3211             & Telecommunication networks             \\
			EFST(R25K)  & 15   & 39483              & 47893              & 25000            & GeoSteiner   generator\\
			EFST(R50K)  & 15   & 79128              & 96293              & 50000            & GeoSteiner generator  \\
			EFST(R100K) & 15   & 158147             & 192190             & 100000           & GeoSteiner   generator \\
			\hline
		\end{tabular}
		
	\end{threeparttable}}
\end{table}

For this set of instances, we adopt the following best STP heuristic algorithms as our reference algorithms.
\begin{itemize}
	\item PUW \cite{pajor2018robust},  a multi-start local search algorithm which ranks first at Formula 1 in the single-thread heuristic track of the 11th DIMACS Competition.
	\item Staynerd \cite{fischetti2017thinning}, a winner of most category of instances in the single-thread heuristic track of the 11th DIMACS Competition.
	\item CIMAT \cite{bonnet2018pace}, the champion of the heuristic track of the PACE Challenge (2018).
	\item MPCH \cite{luipersbeck2013new}, a partition-based algorithm focusing on solving large graphs.
\end{itemize}


For each instance, we perform each of the above algorithms (except MPCH\footnote{Because the code of MPCH is unavailable, we extract its results from \cite{luipersbeck2013new}.}) 10 times independently.
For each algorithm and each instance, we take the best solution obtained in the 10 trials.
 We summarize these results for each dataset in Table \ref{results average gap for each dataset}.
Detailed results of these 113 instances are provided in the full version of the paper.
For each dataset and each algorithm, we indicate the average solution gaps of the algorithm\footnote{For an instance, the gap of a solution $T$ is $\frac{\sum_{e\in E(T)}{c_G(e)} - lb}{lb}\times 100\%$, $lb$ being the best-known lower bound for the instance.} and the number of instances for which the algorithm finds the best solution. In terms of solution gap, our PM algorithm performs better than other algorithms on the largest instances of the GEO, I, and EFST family datasets. It is also remarkable that PM is the best on all I and EFST instances and 16/22 on the GEO instances. Meanwhile, for the relatively smaller instances in this group, the algorithms compete with each other.

{\small
	\begin{table}[h]
		\centering
		\caption{The best solutions found by each algorithms on the set of 113 large instances. The smallest average gaps are marked in bold.}
		\label{results average gap for each dataset}
		\setlength{\tabcolsep}{0.3mm}{
			\resizebox{\columnwidth}{!}{
				\begin{tabular}{ccccccccccc}
					\hline
					\multirow{2}{*}{Dataset} & \multicolumn{5}{c}{ave.gap(\%)}                                          & \multicolumn{5}{c}{\#best} \\
					\cmidrule(lr){2-6} \cmidrule(lr){7-11}
					& MPCH & PUW             & Staynerd        & CIMAT     & PM              &MPCH             & PUW             & Staynerd        & CIMAT    & PM      \\
					\hline
					ES                                           & 0.84 & 0.0082          & \textbf{0.0069} & 0.0987 & 0.0078          & -    & 0   & 3        & 0  & 1  \\
					TSPFST                                       & 0.96 & \textbf{0.0100} & 0.0203          & 0.1068 & 0.0380          & -    & 2   & 2        & 0  & 0  \\
					VLSI                                         & 1.64 & \textbf{0.0292} & 0.3963          & 0.0665 & 0.0730          & -    & 2   & 0        & 1  & 1  \\
					GEO                                          & -    & 0.8136          & 1.0044          & 1.0753 & \textbf{0.7744} & -    & 4   & 0        & 0  & 16 \\
					I                                            & 0.02 & 0.0130          & 0.0154          & 0.0394 & \textbf{0.0107} & -    & 0   & 0        & 0  & 10 \\
					EFST(R25K)                                   & -    & 0.0798          & 0.0347          & 0.2761 & \textbf{0.0101} & -    & 0   & 0        & 0  & 15 \\
					EFST(R50K)                                   & -    & 0.1131          & -               & 0.3387 & \textbf{0.0188} & -    & 0   & 0        & 0  & 15 \\
					EFST(R100K)                                  & -    & 0.1344          & -               & 0.3644 & \textbf{0.0353} & -    & 0   & 0        & 0  & 15\\
					\hline
		\end{tabular}}}
	\end{table}
}


\paragraph{Results on the Set of 30 DIMACS Instances.}
We first test 30 instances that are used in the final ranking phase for the single thread heuristic challenge of the 11th DIMACS Implementation Challenge\footnote{https://dimacs11.zib.de/contest/results/SPGp.1.7200.heur.html}. They are representative instances from the well-known SteinLib, Vienna, Copenhagen, PUCN, and GAPS libraries.

For this set of instances, we compare our results with those of six best performing reference algorithms: AB \cite{althaus2014algorithms}, mozartballs \cite{fischetti2017thinning}, polito \cite{biazzo2012performance}, PUW \cite{pajor2018robust}, scipjack \cite{gamrath2017scip} and Staynerd \cite{fischetti2017thinning}, which participated in the final heuristic track of the 11th DIMACS Challenge. Following the time constraints in the competition results of the DIMACS challenge heuristic track (1 thread), all algorithms are run under a cutoff time of 2 hours.

According to Table \ref{results general graphs}, in terms of finding the best solutions, polito and PM each have 6 solutions that outperform the remaining algorithms. Mozartballs performs best on 'Average' and 'ave. gap(\%)'.  
In particular, a detailed data analysis reveals that PM has the best solution for fnl4461fst-p, alue7080-p, es10000fst01-p, cc3-12p-p, G106ac-p, I064ac-p, which are the largest instances in the group.

\begin{table}[]
	\centering
	\caption{The best solutions found by each algorithms on DIMACS competition instances}
	\label{results general graphs}
	\footnotesize
	\setlength{\tabcolsep}{0.3mm}{
		\begin{tabular}{cccccccc}
			\hline
			Instance       & AB           & mozartballs     & polito          & PUW              & scipjack        & staynerd        & PM                          \\
			\hline
			ave.gap(\%)    & 776.0807     &\textbf{0.3640}  & 3.8920          & 0.4174           & 1.5015          & 0.3737          & 0.6219                      \\
			\#best         & 0            & 0               & \textbf{6}      & 4                & 0               & 1               & \textbf{6}                  \\
			\hline
	\end{tabular}}
\end{table}

\paragraph{The Effectiveness of Different Merging Rules}
We show the average solution gaps obtained by using each rule in Table \ref{results merging}.
 We use the Fredman test to check the statistical significance of the differences in terms of solution quality, which reveals a $p$-value of $0. 00$.
The $p$-value ($< 0.05$) clearly indicates the differences between the variants with the three different merging rules.
The random-edge and min-distance strategies perform better than the random strategy, while min-distance shows a slight advantage over random-edge.

\begin{table}[]
\caption{The average gaps (\%) of PM variants using different merging rules.}
\label{results merging}
\centering

\setlength{\tabcolsep}{1mm}{
	\begin{tabular}{cccc}
		\hline
		Dataset     & random  & random-edge          & min-distance     \\
		\hline
		ES          & 0.02069 & 0.00922          & \textbf{0.00783} \\
		TSPFST      & 0.01832 & \textbf{0.01831}          & 0.03799 \\
		VLSI        & 0.09954 & 0.08684          & \textbf{0.07299} \\
		GEO         & 0.80160 & 0.78055          & \textbf{0.77435} \\
		I           & 0.01144 & 0.01090          & \textbf{0.01074} \\
		EFST(R25K)  & 0.01423 & \textbf{0.00960} & 0.01008          \\
		EFST(R50K)  & 0.02740 & \textbf{0.01868} & 0.01877          \\
		EFST(R100K) & 0.04900 & 0.03646          & \textbf{0.03533} \\
		\hline
\end{tabular}}
\end{table}

\section{Conclusion}

In this paper, we introduced a new partition-and-merge algorithm to effectively solve large-scale STP instances. To ensure the effectiveness of the algorithm, we investigated different merging heuristics, local search algorithms, and new solution storing techniques.

This is the first STP algorithm based on a multi-level merging and search strategy. The algorithm has some attractive features for parallel computing. For example, we can run  local optimizing procedures on different threads in parallel as they are independent.

\section*{Acknowledgements}
We thank the authors of Staynerd, PUW, CIMAT and other solvers in the paper for providing their codes.
We also thank the organizers of the 11th DIMACS challenge and the 3rd PACE competition for hosting the data.
This work is partially funded by the National Natural Science Foundation of China under grant No. 62372093 and the Natural Science Foundation of Sichuan Province of China under grant No. 2023NSFSC1415.

\bibliographystyle{splncs04}
\bibliography{STP-PM-LNCS-FINAL}

\end{document}